\documentclass[nouppercase]{kaia}
\usepackage{amsmath}
\usepackage{cleveref}
\usepackage[labelformat=simple]{subcaption}
\usepackage{float}

\title{XAI based Performance Preserving Adaptive Image Compression for //
Efficient Satellite Communication}

\author{KyungChae Lee}{a}
\affiliation{School of Electrical Engineering, Korea Advanced Institute of Science and Technology, Daejeon, Korea,\\ \{kyungchae.lee\}@kaist.ac.kr}{a}
\begin{document}

\maketitle
\begin{abstract}
	In the era of multinational cooperation, gathering and analyzing the satellite images are getting easier and more important. Typical procedure of the satellite image analysis include transmission of the bulky image data from satellite to the ground producing significant overhead. To reduce the amount of the transmission overhead while making no harm to the analysis result, we propose a novel image compression scheme RDIC in this paper. RDIC is a reasoning based image compression scheme that compresses an image according to the pixel importance score acquired from the analysis model itself. From the experimental results we showed that our RDIC scheme successfully captures the important regions in an image showing high compression rate and low accuracy loss.
\end{abstract}

\section{Introduction}\label{sec:intro}
Satellite image analysis is a crucial task for gathering information world-wide in the era of multinational cooperation. Since most, if not all, satellites have little computational resources satellite image analysis is typically done on the ground where powerful computing stations exist. Thus, analyzing a satellite image starts with image transmission from satellite to the ground station.

The transmission latency is quite large when it comes to the satellite environment since they are very far away from the ground and therefore have very limited communication bandwidth. Therefore, there is a need for an image compression technique in order to mitigate the transmission overhead. However, compressing an image often causes information loss resulting in analysis performance degradation which is not wanted behavior. 

One of the most popular image compression technique is JPEG\cite{jpeg}. JPEG is an image compression standard that has been used in various domains enabling lightweight image compression with acceptable visual image quality. However, since the jpeg compression basically works with filtering out the high frequency color components in the image\cite{jpeg}, color distortion is inevitable after the pipeline of encoding and decoding. This color distortion causes performance degradation on image analysis tools such as deep neural networks. 

In this paper we propose an adaptive image compression technique that can reduce the transmission latency while preserving the accuracy of the analysis tools. This can be done by adaptively choosing the region of interest(RoI) where the analysis tool values the most and compress those regions with high quality and others with low quality. The details of choosing the RoI will be delivered in \Cref{sec:method} We showed that with our proposed algorithm we can significantly reduce the image file size while successfully preserving the analysis accuracy on satellite images.

\section{Related Works}
\subsection{JPEG Image compression for Satellite Imageries}
As mentioned in the \Cref{sec:intro}, jpeg image compression is one of the most popular standard in the domain and there are numerous works that applied jpeg compression on satellite images\cite{jpegeval}. Work done by Tada et. al.\cite{jpegeval} evaluated the effect of the jpeg compression with power spectrum comparison showing the degree of image distortion after the compression. What is not dealt with tada's work is that the image compression may cause the deep neural network malfunction even though the compressed image looks fine with human eyes. Researches like \cite{jpegimpact1, jpegimpact2} report that jpeg compression does affect the performance of the deep neural network. We also conducted a preliminary experiment that shows the impact of jpeg compression on Faster R-CNN object detection network for satellite images in \Cref{sec:experiment}. Thus, it is clear that compressing a satellite image should be dealt carefully in order to get the maximum analyzing performance on the ground. 

\subsection{Image Compression with Accuracy Consideration}
There are some works that deals with the image compression considering the outcome of the analysis tools on the receiver side\cite{dre_pre, dre_pre2,dre_pre3,dre_pre4,edge_assist}. Both works are targetting the different domain but the basic idea is similar. Work \cite{edge_assist} uses a image compression for efficient image transmission between mobile client and the edge server when offloading the compute intensive image object detection task from client to the server. Here, they propose a Dynamic Region of Interest for adaptive image compression where the object detection result for the previous frame is used for determining the important area of the current frame. The selected RoI is then compressed with higher quality(less compression), and the rest of the image area are compressed with lower quality(more compression). By doing so, \cite{edge_assist} achieves real-time image offloading in edge assisted augmented reality(AR) service. 

Work \cite{dre_pre} and others \cite{ dre_pre2,dre_pre3,dre_pre4} are more related to our work that those directly aim the same domain we are dealing with: Satellite image compression without performance degradation. Paper \cite{dre_pre} proposes a fuzzy c-means image segmentation and adaptive image compression according to the segmentation result which in turn compresses the background more and important objects less. 

However, works like \cite{dre_pre, dre_pre2,dre_pre3,dre_pre4} simply focuses on first incestigatin the image and do not directly consider the structure or the characteristics of the analysis tools on the ground. This inconsistency of image interpretation between land and satellite would result in performance degradation. On the other hand, work done by \cite{edge_assist} uses the previous result from the analysis tool, so it directly considers the analysis tool when compressing an image. Nonetheless, algorithm of \cite{edge_assist} is very hard to be applied to satellite imageries since satellite images typically captures different places and even if the time series data can be produced, huge transmission overhead makes data from last time step less valuable. 

\begin{figure}[!bp]
    \centering
    \begin{subfigure}{0.45\linewidth}{
        \centering
        \includegraphics[width=\linewidth]{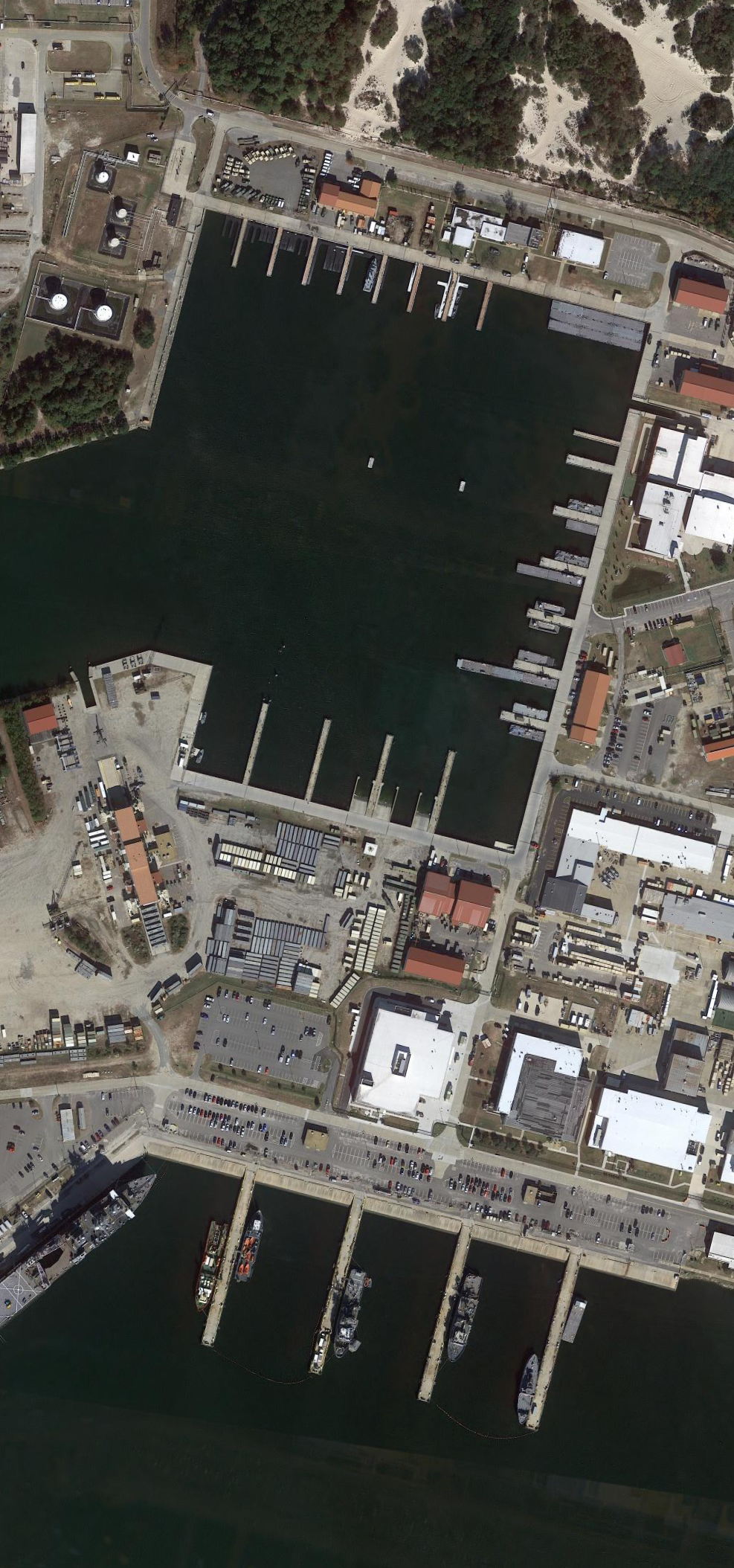}
        \caption{Input image}
    }
    \end{subfigure}
    \begin{subfigure}{0.45\linewidth}{
        \centering
        \includegraphics[width=\linewidth]{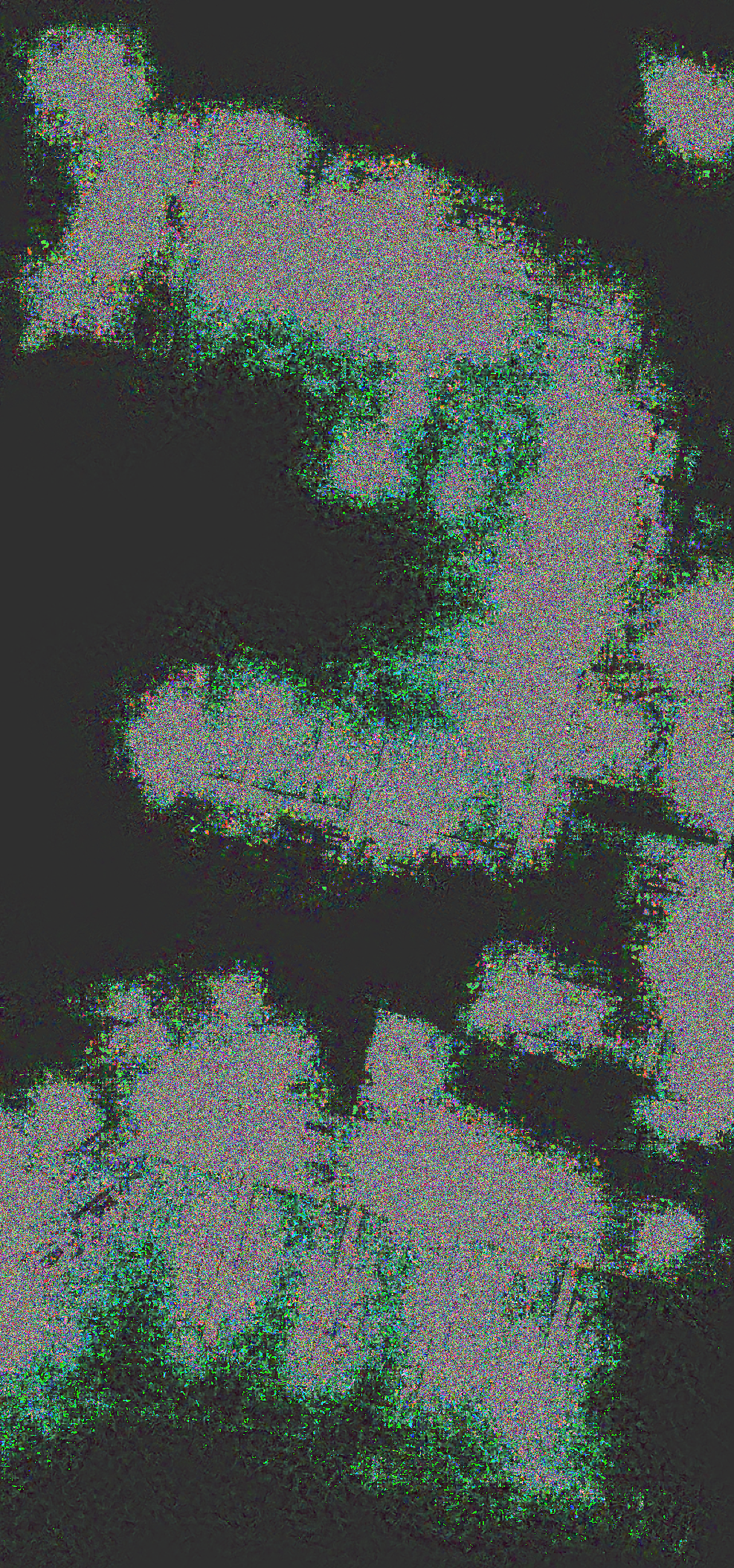}
        \caption{Epsilon lrp result}
    }
    \end{subfigure}
	\caption{Example of applying epsilon lrp model explanation to the satellite image input and object dection model}
	\label{fig:elrp}
\end{figure}

\section{Methodology}\label{sec:method}
In this paper, we propose a novel image compression scheme, reasoning based dynamic image compression(RDIC), which makes use of the layer-wise relevance propagation\cite{elrp} which is one of the explainable AI techniques. Layer-wise relevance propagation works by backpropagating the neural network result and it's relevance score as in the equation \ref{eqn:elrp1} and point the salient part of the input image where the model got the most valuable information getting the result. 

\begin{equation}\label{eqn:elrp1}
R^{(l)}_i = \sum_{j} {{z_{ij}}\over{\sum_{i'} z_{i'j} + \epsilon sign(\sum_{i'} z_{i'j})}} R^{(l+1)}_j
\end{equation}

An example of the result of applying epsilon lrp, layer-wise relevance propagation, to the object detection model SCRDet\cite{r2cnn} is shown in the Figure. \ref{fig:elrp}. As can be seen in the figure salient part containing target objects(ship, harbor, cars, etc...) are highlighted. Instead of highlighting the whole foreground objects by using epsilon lrp we could highlight the image region that is needed by the model for image analysis.

\begin{figure}[!tp]
    \centering
    \begin{subfigure}{0.45\linewidth}{
        \centering
        \includegraphics[width=\linewidth]{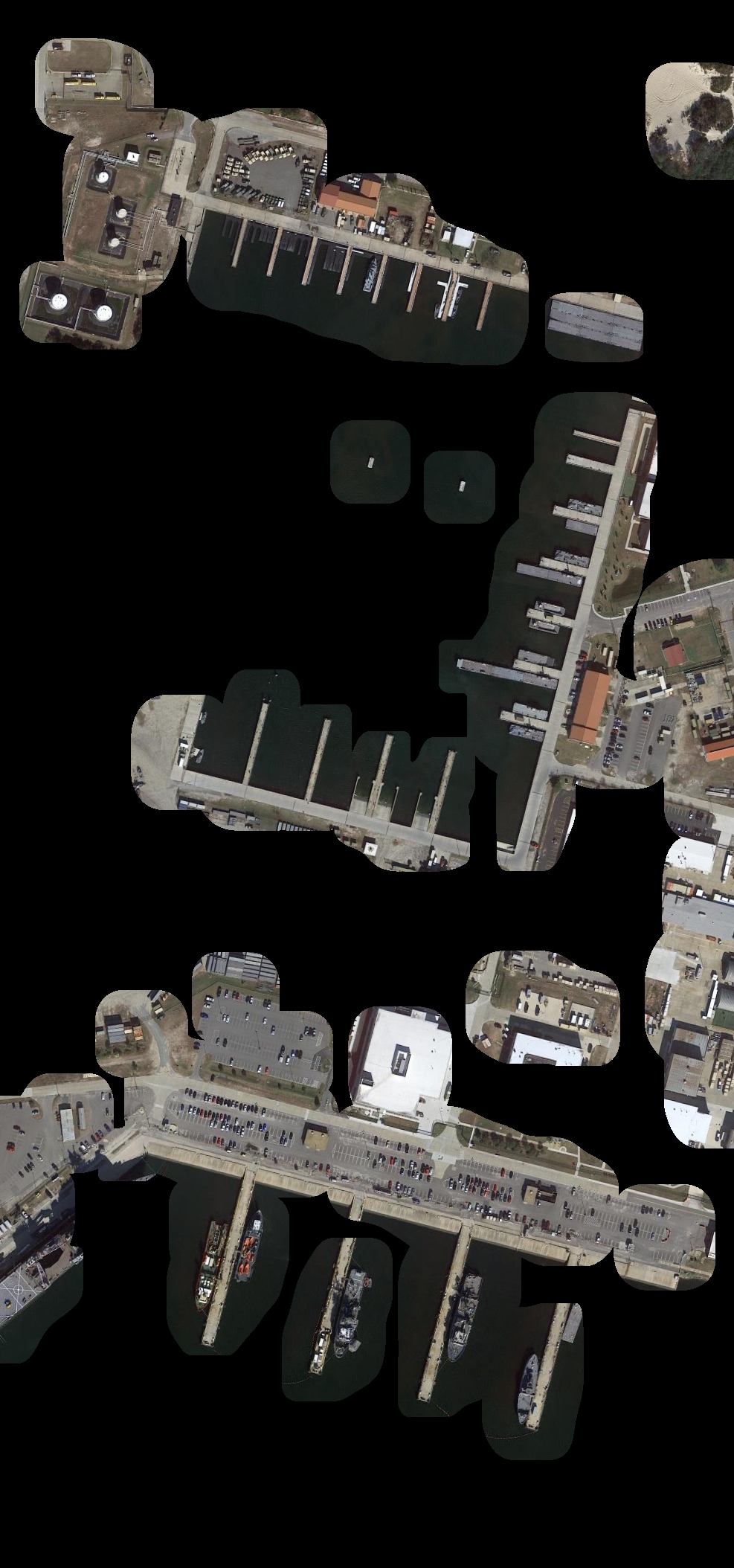}
        \caption{Salient region}
    }
    \end{subfigure}
    \begin{subfigure}{0.45\linewidth}{
        \centering
        \includegraphics[width=\linewidth]{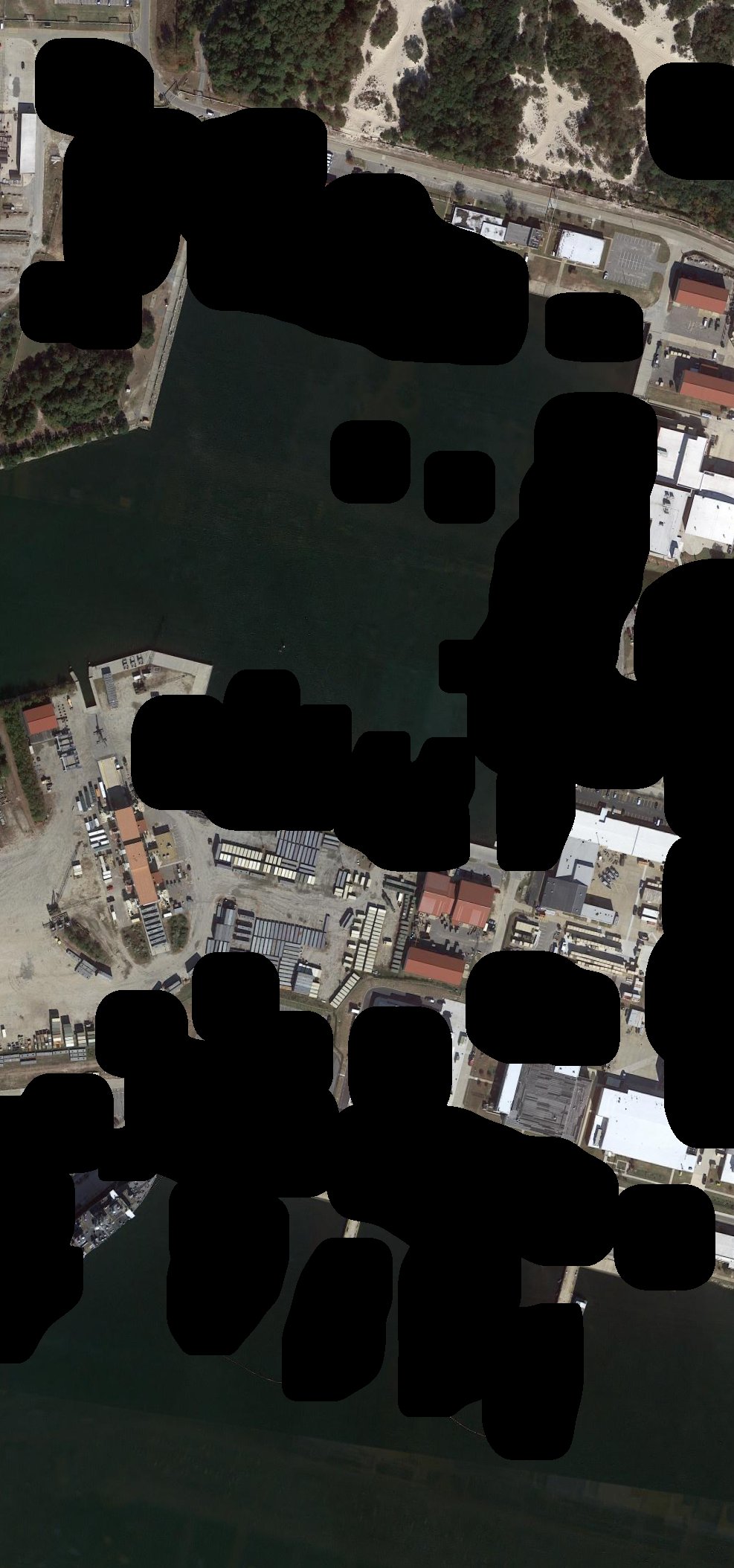}
        \caption{Background}
    }
    \end{subfigure}
	\caption{Example of calculated final RoI mask}
	\label{fig:masks}
\end{figure}

\begin{figure*}[!htbp]
    \centering
    \includegraphics[width=\linewidth]{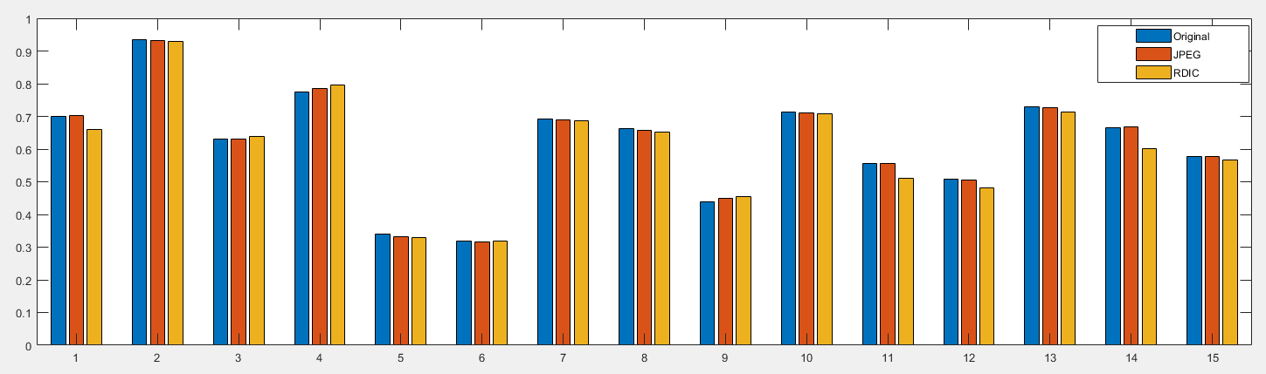}
    \caption{Experimental results for comparing the mAP score for each test cases: Original Dataset(Blue), JPEG compressed dataset(Orange), RDIC compressed dataset(Yellow). Class number(1-16) stands for classes ['roundabout', 'tennis-court', 'storage-tank', 'soccer-ball-field', 'small-vehicle', 'ship', 'plane', 'large-vehicle', 'helicopter', 'harbor', 'ground-track-field', 'bridge', 'basketball-court', 'baseball-diamond', 'mAP']}
	\label{fig:results}
\end{figure*}
From the result of the epsilon lrp relevance propagation, we now calculate the region of interest(RoI) which will act as a mask determining the compression quality. The outcome of the epsilon lrp is a bitmap which indicates the pixel-wise relevance score within the range of negative infinity and positive infinity. This unboundedness makes the calculation of the importance mask difficult with raw outcome of the elrp. Thus, we first took the absolute value of the outcome and then noramlized with the mean value of the outcome. From the normalized outcome values, we then create a mask M which indicates the pixel area where the normalized relevance value is bigger than 0 as shown in the equation \ref{eqn:mask}. However, as we can see from the Fig. \ref{fig:elrp}-(b), the outcome of the elrp is basically very noisy and therefore the resulting mask is also very noisy. The noisiness of the mask can significantly degrade the performance of the target object detection model so we implemented a dilation operation, one of the conventional CV techniques, for acquiring smooth RoI masks. The example of the final RoI mask can be seen in the Fig. \ref{fig:masks}. As can be seen in the figure, salient areas containing the target objects are successfully highlighted. Furthermore, in the background region we can see that salient objects are also included but the category of the salient objects does not belong to the target category.

\begin{equation}\label{eqn:mask}
M(i,j) = \begin{cases}
1 &\text{$abs(elrp(I)(i,j)) >= mean(abs(elrp(I)))$}\\
0 &\text{else, I: Input image}
\end{cases}
\end{equation}

After the calculation of the RoI for determining the compression criteria, we conducted the dynamic image compression where we compress the RoI region with high quality and background region with low quality. Here, the quality of the compression follows the quality definition of the computer vision OpenCV\cite{opencv}.

\section{Experimental Results}\label{sec:experiment}
For the evaluation of our proposed image compression scheme we have conducted an experiment comparing the mean average precision performance of the object detection model on datasets compressed with different methods. 

For the dataset we used a DOTA dataset\cite{dota} which is an open dataset consisting of numerous images taken by a plane that is similar to the satellite imageries.

For the target object detection model we chose faster r-cnn model following the paper \cite{r2cnn}. 

We compared the evaluation result of the faster r-cnn model on the dataset first, and then we compressed the original dataset with two different methods: Original JPEG compression, and proposed reasoning based dynamic image compression(RDIC). Here, original jpeg compression is done with the quality of 100, and RDIC consist of two different quality 100, 50 each for RoI and BG regions. We then compared both the mAP score and the total file size of the dataset which can be seen in the Fig. \ref{fig:results} and the Table. \ref{tab:results}.

\begin{table}[ht]
\begin{center}
\begin{tabular}{|c|c|c|c|}
\hline
 & Original & JPEG& RDIC\\
\hline\hline
File Size(MB) & 3324 & 1671 & 942\\
\hline
mAP(\%) & 57.75 & 57.75 & 56.54\\

\hline
\end{tabular}
\end{center}
\caption{Example of caption for table.}
\label{tab:results}
\end{table}

The figure Fig. \ref{fig:results} shows the average precision of each classes. As we can see from the figure, the performance of the object detection model is mostly preserved after the compression. Interesting part is that in case of classes like soccer-ball-field and large-vehicle, jpeg and RDIC compressed version of ours resulted in a better precision score. Except for this unexpected outcome, we can see that the average precision gets lower when applied compression to the dataset. However, if we look into the file size analysis in the Table. \ref{tab:results}, we can see that compared to the original dataset, JPEG compression provides identical performance while the size of the dataset is reduced to 50.27 percent of the original dataset. Our proposed RDIC loses about 1.21 percent point of the accuracy while reducing the filesize into 942 Mega Bytes which is 56.4 percent of the JPEG compressed dataset, and 27.9 percent of the original dataset. This significant reduction of the filesize allows the satellite image transmission to be about four times faster than usual with only a 1.2 percent point loss of the detection model accuracy.

\section{Conclusion}\label{conclusion}
Satellite imageries are big in their size which causes a huge transmission latency hindering the fast and easy analysis of the image. In this paper we propose a novel image compression scheme based on the model reasoning that allows us to compress the satellite image with minimum accuracy loss and high compression rate. Our scheme starts from analyzing the target model by relevance propagation for RoI searching. According to the RoI we then conducted a dynamic image compression which will compress the important part of an image with high quality and others with high compression rate. The evaluation results show that our scheme successfully capture the important region in the image according to the model we use. Since the epsilon lrp method and other techniques we used is not bound to a single object detection model, our scheme is also easy to apply on various other applications and neural network models.
% \section{Some Examples}
% \subsection{Equation}
% \begin{equation}
% \ell = -\sum^M_{c=1} y_{o,c} \log (p_{o,c})	
% \end{equation}

% \subsection{Figures}
% \begin{figure}[ht]
% \begin{center}
% \end{center}
%   \caption{Example of caption for figure.}
% \end{figure}

% \subsection{Table}
% \begin{table}[ht]
% \begin{center}
% \begin{tabular}{|l|c|}
% \hline
% A & B \\
% \hline\hline
% C & D \\
% E & F \\
% G & H\\
% \hline
% \end{tabular}
% \end{center}
% \caption{Example of caption for table.}
% \end{table}

\bibliographystyle{plain}
\bibliography{references}

\end{document}